\documentstyle[10pt,psfig]{article}
\renewcommand{\author}[2]{\begin{center}
{\sc #1}\\
{#2}
\end{center}}
\renewcommand{\title}[1]{\begin{center}
{\Large {\bf #1}}
\end{center}}

\begin{document}

\title{Numerical simulations of relativistic wind accretion
on to black holes using Godunov-type methods
\vspace*{5mm}}

\author{Jos\'e A. Font}
\noindent
Max-Planck-Institut f\"ur Astrophysik \\
Karl-Schwarzshild-Str. 1, D-85740 Garching, Germany 

\noindent
e-mail: font@mpa-garching.mpg.de


\author{Jos\'e M$^{\underline{\mbox{a}}}$. Ib\'a\~nez}
\noindent
Departamento de Astronom\'{\i}a y Astrof\'{\i}sica,
Universidad de Valencia \\
46100 Burjassot (Valencia), Spain

\noindent
e-mail: ibanez@godunov.daa.uv.es

\author{Philippos Papadopoulos}
\noindent
School of Computer Science and Mathematics, 
University of Portsmouth \\
PO1 2EG, Portsmouth, United Kingdom

\noindent
e-mail: Philippos.Papadopoulos@port.ac.uk

\vspace{5mm}

\begin{abstract}
We have studied numerically the so-called Bondi-Hoyle
(wind) accretion on to a rotating (Kerr) black hole in 
general relativity. We have used the Kerr-Schild form of the Kerr metric,
free of coordinate singularities at the black hole horizon.
The `test-fluid' approximation has been adopted, assuming no dynamical
evolution of the gravitational field. We have used a recent formulation of the 
general relativistic hydrodynamic equations which casts them into a first-order 
hyperbolic system of conservation laws. Our studies have been performed using 
a Godunov-type scheme based on Marquina's flux-formula.

We find that regardless of the value of the black hole spin the final accretion 
pattern is always stable, leading to constant accretion rates of mass and 
momentum. The flow is characterized by a strong tail shock 
which is increasingly wrapped around the central black hole as the
hole angular momentum increases.  The rotation induced asymmetry in
the pressure field implies that besides the well known drag, the black
hole will experience also a {\em lift} normal to the flow direction.

\end{abstract}


\section{Introduction}

The term ``wind" or hydrodynamic accretion refers to the capture of matter by a
moving object under the effect of the underlying gravitational field.
The canonical astrophysical scenario in which matter is accreted in such a 
non-spherical way was suggested originally by Bondi and Hoyle~\cite{bondi44}, 
who studied, using Newtonian gravity, the accretion 
on to a gravitating point mass moving with constant velocity through a non-relativistic 
gas of uniform density. Such process applies to describe
mass transfer and accretion in compact X-ray binaries, in particular
in the case in which the donor (giant) star lies inside its Roche lobe and
loses mass via a stellar wind. This wind impacts on the orbiting
compact star forming a bow-shaped shock front around it.

The problem was first numerically investigated in the early 70's. Since then, 
contributions of a large number of authors using highly developed Godunov-type 
methods extended the simplified analytic models (see, e.g.,~\cite{ruffert94,
benensohn97} and references there in). These Newtonian investigations 
helped develop a thorough understanding of the hydrodynamic accretion
scenario, in its fully three-dimensional character, revealing the
formation of accretion disks and the appearance of non-trivial phenomena
such as shock waves or {\it flip-flop} instabilities. 

We have recently considered hydrodynamic accretion on to a moving black
hole using relativistic gravity and the ``test fluid" 
approximation~\cite{font98a,font98b,font98,font99}.
We present here a brief summary of the methodology and results of
such simulations. We integrate the general relativistic hydrodynamic
equations in the fixed background of the Kerr spacetime 
(including its non-rotating Schwarzschild limit) and
neglect the self-gravity of the fluid as well as non-adiabatic processes 
such as viscosity or radiative transfer.
In the black hole case the matter flows ultimately across the event horizon and
becomes causally disconnected of distant observers .
Near that region the problem is intrinsically relativistic 
and the gravitational accelerations significantly
deviate from the Newtonian values.

\section{Equations}

The general relativistic hydrodynamic equations can be cast as a 
first-order flux-conservative system describing the conservation of mass, 
momentum and energy. Formulations of this sort are given, e.g. in~\cite{banyuls97,
papadopoulos99}.  In this work we follow the approach laid out in~\cite{banyuls97}
for a perfect fluid stress-energy tensor $T^{\mu\nu}$. 
The system of equation then reads:
\begin{equation}
\frac{1}{\sqrt{-g}} \left(
\frac {\partial \sqrt{\gamma}{\bf u}}
{\partial x^{0}} +
\frac {\partial \sqrt{-g}{\bf f}^{i}}
{\partial x^{i}} \right)
 = {\bf s}
\label{F}
\end{equation}
\noindent
($x^0=t$; $x^i$ spatial coordinates, $i=1,2,3$)
where ${\bf u}\equiv {\bf u}({\bf w})$ are the evolved quantities,
${\bf u} = (D, S_j, \tau)$
and ${\bf f}^{i}$ are the fluxes
\begin{equation}
{\bf f}^{i}  =   \left(D \left(v^{i}-\frac{\beta^i}{\alpha}\right),
 S_j \left(v^{i}-\frac{\beta^i}{\alpha}\right) + p \delta^i_j,
\tau \left(v^{i}-\frac{\beta^i}{\alpha}\right)+ p v^{i} \right),
\end{equation}
\noindent
$v^i$ being the 3-velocity and $p$ the pressure.
The corresponding sources ${\bf s}$ are given by
\begin{eqnarray}
{\bf s} =  \left(0,
T^{\mu \nu} \left(
\frac {\partial g_{\nu j}}{\partial x^{\mu}} -
\Gamma^{\delta}_{\nu \mu} g_{\delta j} \right),
\alpha  \left(T^{\mu 0} \frac {\partial {\rm ln} \alpha}{\partial x^{\mu}} -
T^{\mu \nu} \Gamma^0_{\nu \mu} \right)
                     \right).
\end{eqnarray}
\noindent
We note the presence of geometric terms in the fluxes and sources which 
appear as the local conservation laws of the density current and stress-energy
are expressed in terms of partial derivatives. These terms are the
lapse function $\alpha$, the shift vector $\beta^i$ and the connection coefficients
$\Gamma^{\delta}_{\nu \mu}$ of the 3+1 spacetime metric
\begin{equation}
ds^{2} \equiv g_{\mu\nu}dx^{\mu}dx^{\nu} = -(\alpha^{2}-\beta_{i}\beta^{i}) dt^{2}+
2 \beta_{i} dx^{i} dt + \gamma_{ij} dx^{i}dx^{j}
\end{equation}
\noindent
Additionally $g\equiv \det(g_{\mu\nu})$ is such that
$\sqrt{-g} = \alpha\sqrt{\gamma}$ and $\gamma\equiv \det(\gamma_{ij})$.

The vector ${\bf w}$, representing the primitive variables, is
given by ${\bf w} = (\rho, v_{i}, \varepsilon)$ where $\rho$ is
the density and $\varepsilon$ the specific internal energy. The evolved
quantities are defined in terms of the primitive variables as
$D=\rho W$, $S_j = \rho h W^2 v_j$ and $\tau=\rho h W^2 - p - D$,
$W$ being the Lorentz factor $W = (1-v^2)^{-1/2}$, 
with $v^2=\gamma_{ij}v^i v^j$, and $h$ the specific
enthalpy, $h=1+\varepsilon+p/\rho$. A perfect fluid equation of state
$p=(\Gamma - 1) \rho \varepsilon$, $\Gamma$ being the
constant adiabatic index, closes the system.

In our computations we specialize the above expressions to the Kerr 
line element which describes the exterior geometry of a rotating black
hole. We use the Kerr-Schild form of the Kerr metric, which is free
of coordinate singularities at the black hole horizon. Computations
using the more standard Boyer-Lindquist (singular) form of the metric
are presented in~\cite{font99}. Pertinent technical details concerning the 
specific form of these metrics are given in~\cite{papadopoulos98}.

\section{Numerical scheme}

Our hydrodynamical code performs the numerical integration of system
(\ref{F}) using a Godunov-type method.  The time update from $t^n$ to 
$t^{n+1}$ proceeds according to the following algorithm in conservation
form:
\begin{eqnarray}
    {\bf u}_{i,j}^{n+1} = {\bf u}_{i,j}^{n}
     - \frac{\Delta t}{\Delta x^k}
    (\widehat{{\bf f}}_{i+1/2,j}-\widehat{{\bf f}}_{i-1/2,j})
 +  \Delta t \,\, {\bf s}_{i,j} \, ,
\end{eqnarray}
\noindent
improved with the use of (second-order) conservative Runge-Kutta sub-steps 
to gain accuracy in time~\cite{shu89}. The numerical fluxes are computed by 
means of Marquina's flux-formula~\cite{donat96}.
After the update of the conserved quantities the primitive variables are 
computed via a root-finding procedure. 

The flux-formula makes use of the complete characteristic information
of system (\ref{F}), eigenvalues (characteristic speeds) and right and
left eigenvectors. Generic expressions are collected in~\cite{ibanez99}.

The state variables, ${\bf u}$, must be computed
(reconstructed) at the left and right sides of a given interface, out of
the cell-centered quantities, prior to compute the numerical fluxes.
In relativistic hydrodynamics one has the freedom to reconstruct either
${\bf w}$ (primitive variables) or ${\bf u}$ (evolved variables). For
efficiency and accuracy considerations we reconstruct the first set,
from which the remaining variables are obtained algebraically.
The code uses slope-limiter methods to construct second-order TVD schemes 
by means of monotonic piecewise linear reconstructions of the cell-centered 
quantities. We use the standard minmod slope which provides the desired 
second-order accuracy for smooth solutions, while still satisfying the TVD 
property.

\section{Results}

\begin{figure}
\centerline{\psfig{figure=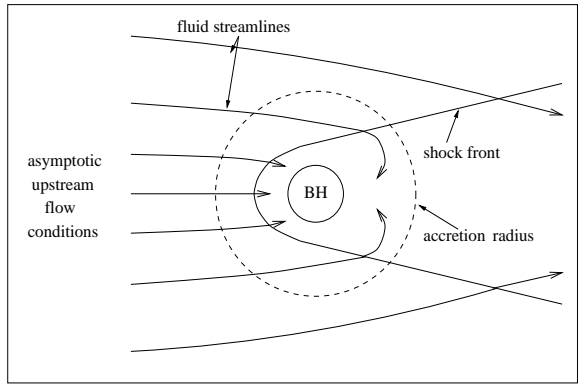,width=3.0in,height=2.0in}}
\caption{{ \protect \small Schematic representation of stationary
supersonic wind accretion. The shock may be detached as in the figure 
(bow shock) or attached to the rear part (tail shock),
depending on the flow asymptotic conditions ($v_{\infty}$
and $c_{s_{\infty}}$) and thermodynamics of the gas ($\rho_{\infty}$ and 
$\Gamma$). 
}}
\label{fig1}
\end{figure}

The classical solution for an asymptotically uniform wind of presureless
gas past a compact source (modeled analytically by a point mass)
was obtained by~\cite{bondi44}. In this solution the material is 
focussed at the rear part of the object as a result of the gravitational pull. 
For a pressureless gas, the density at this symmetry line could reach an infinite
value and matter would flow on to the hole along this accretion line. However,
when pressure is included in the model, a cylindrical shock forms
around this line and the accretion proceeds along an accretion column of high
density and pressure shocked material.
The predicted final accretion pattern consists of a stationary conical shock with 
the material inside the accretion radius being captured by the central
object. An schematic representation of this solution is depicted in
Fig.~\ref{fig1}.

\begin{figure}
\centerline{\psfig{figure=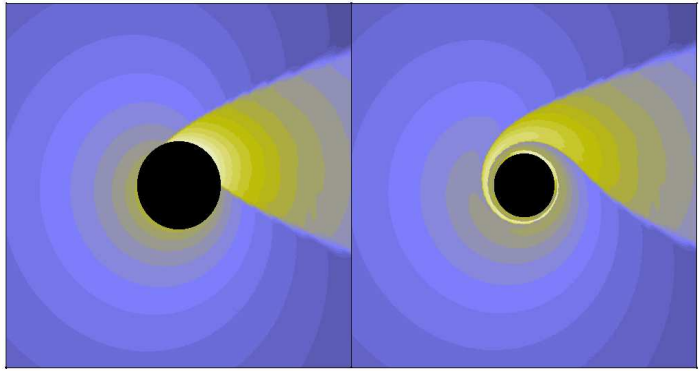,width=4.5in,height=2.3727in}}
\caption{{ \protect \small
Relativistic wind accretion on to a rapidly rotating
Kerr black hole ($a=0.999M$, the hole spin is counter-clock wise)
in Kerr-Schild coordinates (left panel). Initial model parameters:
$v_{\infty}=0.5$, $c_{s_{\infty}}=0.1$ and $\Gamma=5/3$.  Isocontours
of the logarithm of the density are plotted at the final stationary
time $t=500M$. The right panel shows how the flow solution
looks like when transformed to Boyer-Lindquist coordinates.
The shock appears here totally wrapped around the horizon of
the black hole. The box is $12M$ units long. The simulation
employed a $(r,\phi)$-grid of $200\times 160$ zones.
}}
\label{fig2}
\end{figure}

A numerical evolution of relativistic wind accretion past a 
rapidly-rotating Kerr black hole ($a=0.999M$, $a$ specific angular
momentum, $M$ black hole mass) is depicted in Fig.~\ref{fig2} 
(left panel). This simulation shows the steady-state pattern in the equatorial 
plane of the black hole. The tail shock appears stable to
tangential oscillations, in contrast to Newtonian simulations with
tiny accretors (see, e.g.,~\cite{benensohn97} and references there
in; see~\cite{font98b} for a related discussion). 
The accretion rates of mass and linear and angular momentum
also show a stationary behavior (see, e.g.,~\cite{font98b,font99}).
As opposed to the non-rotating black hole, in the rotating case the shock
becomes wrapped around the central accretor, the effect being more
pronounced as the black hole angular momentum $a$ increases. 
The inner boundary of the domain is located at $r=1.0M$
({\it inside} the event horizon which, for this model, is at $1.04M$)
which is only possible with the adopted regular coordinate system.
The flow morphology shows smooth behavior when crossing the
horizon, all matter fields being regular there.

The enhancement of the pressure in the post-shock zone is responsible
for the ``drag'' force experienced by the accretor.  
The rotating black hole redistributes the high
pressure area, with non-trivial effects on the nature of the drag
force. The pressure enhancement is
predominantly on the counter-rotating side. We
observe a pressure difference of almost two orders of magnitude, along
the axis normal to the asymptotic flow direction.  The implication of
this asymmetry is that a rotating hole moving across the interstellar
medium (or accreting from a wind), will experience, on top of the drag
force, a ``lift'' force, normal to its direction of motion (to the
wind direction). Although different in origin this feature bears a 
superficial resemblance with the Magnus effect of classical fluid dynamics.

The right panel of Fig.~\ref{fig2} shows how the accretion pattern would 
look like were the computation performed using the more common (though singular) 
Boyer-Lindquist coordinates. The transformation induces a noticeable
wrapping of the shock around the central hole. The shock would wrap
infinitely many times before reaching the horizon. As a result, the computation in 
these coordinates would be much more challenging than in Kerr-Schild
coordinates, particularly near the horizon. 
Since the last stable orbit approaches closely the horizon in
the case of maximal rotation, the interesting scenario of co-rotating
extreme Kerr accretion would be severely affected by the strong
gradients which develop in the strong-field region.
This will most certainly affect the accuracy and,
potentially, also the stability of numerical codes.

\paragraph{\it Acknowledgments:}J.A.F. acknowledges financial
support from a TMR fellowship of the European Union
(contract nr. ERBFMBICT971902).



\begin{thebibliography}{99}                                             %

\bibitem{banyuls97}
Banyuls F, Font JA,
Ib{\'a}{\~n}ez JM$^{\underline{\mbox{a}}}$,
Mart\'{\i} JM$^{\underline{\mbox{a}}}$, and Miralles JA (1997)
Numerical 3+1 general relativistic hydrodynamics: A local characteristic 
approach. 
{\it ApJ}, {\bf 476}: 221.

\bibitem{benensohn97}
Benensohn JS, Lamb DQ, and Taam RE (1997)
Hydrodynamical studies of wind accretion onto compact objects: 
Two-dimensional calculations,
{\it ApJ}, {\bf 478}: 723.

\bibitem{bondi44}
Bondi H, and Hoyle F (1944) 
On the mechanism of accretion by stars,
{\it MNRAS}, {\bf 104}: 273.

\bibitem{donat96}
Donat R, and Marquina A (1996)
Capturing shock reflections: an improved flux formula,
{\it J. Comput. Phys.}, {\bf 125}: 42.

\bibitem{font98a}
Font JA, and Ib{\'a}{\~n}ez JM$^{\underline{\mbox{a}}}$ (1998)
A numerical study of relativistic Bondi-Hoyle accretion on to a moving 
black hole: Axisymmetric computations in a Schwarzschild background,
{\it ApJ}, {\bf 494}: 297.

\bibitem{font98b}
Font JA, and Ib{\'a}{\~n}ez JM$^{\underline{\mbox{a}}}$ (1998)
Non-axisymmetric relativistic Bondi-Hoyle accretion on to a Schwarzschild 
black hole,
{\it MNRAS}, {\bf 298}: 835.

\bibitem{font98}
Font JA, Ib{\'a}{\~n}ez JM$^{\underline{\mbox{a}}}$, and
Papadopoulos P (1998)
A horizon-adapted approach to the study of relativistic accretion flows 
on to rotating black holes,
{\it ApJ}, {\bf 507}: L67.

\bibitem{font99}
Font JA, Ib{\'a}{\~n}ez JM$^{\underline{\mbox{a}}}$, and
Papadopoulos P (1999)
Non-axisymmetric relativistic Bondi-Hoyle accretion on to a Kerr black 
hole,
{\it MNRAS}, {\bf 305}: 920.

\bibitem{ibanez99}
Ib{\'a}{\~n}ez JM$^{\underline{\mbox{a}}}$,
Aloy MA, Font JA, Mart\'{\i} JM$^{\underline{\mbox{a}}}$,
Miralles JA, and Pons JA (1999)
Riemann solvers in general relativistic hydrodynamics,
this volume.

\bibitem{papadopoulos98}
Papadopoulos P, and Font JA (1998)
Relativistic hydrodynamics around black holes and horizon adapted 
coordinate systems,
{\it Phys. Rev. D}, {\bf 58}: 024005.

\bibitem{papadopoulos99}
Papadopoulos P, and Font JA (1999)
Relativistic hydrodynamics on spacelike and null surfaces: Formalism 
and computations of spherically symmetric spacetimes,
{\it Phys. Rev. D}, in press (gr-qc/9902018).

\bibitem{ruffert94}
Ruffert M, and Arnett D (1994)
Three-dimensional hydrodynamic Bondi-Hoyle accretion. II. 
Homogeneous medium at Mach 3 with $\gamma=5/3$,
{\it ApJ}, {\bf 427}: 342.

\bibitem{shu89}
Shu CW, and Osher S (1989)
Efficient implementation of essentially non-oscillatory shock-capturing
schemes. II,
{\it J. Comput. Phys.}, {\bf 83}: 32.

\end{thebibliography}
\end{document}